# On the Imminent Regional Seismic Activity Forecasting Using INTERMAGNET and Sun-Moon Tide Code Data

Strachimir Cht. Mavrodiev[1], Lazo Pekevski[2], Giorgi Kikuashvili[3], Emil Botev[4], Petar Getsov[5], Garo Mardirossian[5], Georgi Sotirov[5], Dimitar Teodossiev[5]

[1]Institute for Nuclear Research and Nuclear Energy, Bulgarian Academy of Sciences, Sofia, Bulgaria
[2]University "Sts. Cyril and Methodius", Seismological Observatory, Skopje, Macedonia
[3]Ilya State University, Tbilisi, Georgia
[4]National Institute for Geophysics, Geography and Geodesy, Bulgarian Academy of Sciences, Sofia, Bulgaria
[5]Space Research and Technology Institute, Bulgarian Academy of Sciences, Sofia, Bulgaria
Email: schtmavr@yahoo.com

## Abstract

In this paper we present an approach for forecasting the imminent regional seismic activity by using geomagnetic data and Earth tide data. The time periods of seismic activity are the time periods around the Sun-Moon extreme of the diurnal average value of the tide vector module. For analyzing the geomagnetic data behaviour we use diurnal standard deviation of geomagnetic vector components F (North, East, Down) for calculating the time variance GeomagSignal. The Sun storm influence is avoided by using data for daily A-indexes (published by NOAA). The precursor signal for forecasting the incoming regional seismic activity is a simple function of the present and previous day GeomagSignal and A-indexes values. The reliability of the geomagnetic "when, regional" precursor is demonstrated by using statistical analysis of day difference between the times of "predicted" and occurred earthquakes. The base of the analysis is a natural hypothesis that the "predicted" earthquake is the one whose surface energy density in the monitoring point is bigger than the energy densities of all occurred earthquakes in the same period and region. The reliability of the approach was tested using the INTERMAGNET stations data located in Bulgaria, Panagurishte, PAG (Jan 1, 2008-Jan 29, 2014), Romania, Surlari, SUA (Jan 1, 2008-Jan 27, 2014), Italy, L'Aquila, AQU (Jan 1, 2008-May 30, 2013) in the time of EU IRSES BlackSeaHazNet (2011-2014) project. The steps of program for solving the "when, where and how" earthquake prediction problem are shortly described.

## Keywords

**Earthquakes, Geomagnetism, Earth Sun-Moon Tide, Regional Seismic Activity, Forecasting**

## 1. Introduction

The problem of "when, where and how" earthquake prediction cannot be solved only on the basis of geodetic data [1]-[6]. The possible triggering of the earthquakes by the tidal has been investigated for a long period of time [7]-[14]. The conclusion that the earthquake's time is correlated with the tidal extremes is not unique, because in some of the extremes there are no earthquakes. The inclusion of additional information in the analysis, for example the Earth's electrical currents' signals, permits one to estimate the most probable time of incoming earthquakes [15]-[17].

As more accurate space and time measuring set for the Earth's crust condition parameters, including in the monitoring of the measurements of the electromagnetic fields under, on and over the Earth's surface, the temperature distribution and other possible precursors can be useful for the study of the "when, where and how" predictions of earthquakes. For example, in the papers [18]-[20], the possibility for short-term earthquake prediction in Greece by seismic electric signals was investigated.

The atmospheric and ionosphere electromagnetic phenomena associated with earthquakes were analyzed in many books and papers [21]-[26] and the future direction of the investigation related to earthquake prediction was proposed, as well as its practical application to some events. The papers [27] [28] concern the observations of electromagnetic radiation in the LF and VLF ranges related to the occurrence of an earthquake.

The paper [29] presents the results of the complex investigation of the variations of crust electrical resistivity as a function of tidal deformations on the basis of an extremely low-frequency radio station, which could be more promising for increasing the reliability of electromagnetic-based earthquake predictions. In the papers [30] [31] the electromagnetic anomalies in a wide range of radio frequencies from ULF, VLF to VHF, which had been observed before some destructive earthquakes in continental Greece, are presented.

The impressive results of the modified VAN method are presented in papers and via web sites [15]-[17] [32] [33], where the appropriate measuring of electric Earth signals is analyzed. Their analysis demonstrates that the direction to the epicentre of incoming earthquakes can be estimated and the time is defined from the next extreme of tidal potential. Some possible geophysical models of the phenomena are proposed and the prediction of the future magnitude is analyzed.

On the web site [34] and in the papers cited there, the results of electropotential monitoring, based on the especially constructed electrometer and the appropriate temporal data acquisition system, are presented.

One has to mention the satellite possibilities for monitoring the radiation activity of the Earth's surface for discovering the anomalies, which should be earthquake precursors [35].

The analyses of the data from satellite monitoring for the ionosphere and the Earth's radiation belt parameters also give evidence for anomalies which can be interpreted as earthquake precursors. The information concerning the latest results of the developing of earthquake precursor research can be found on the conference sites [36] [37].

The data for the connection between an incoming earthquake and meteorology effects like quasi-stationary earthquake clouds can be seen on the website [38]. The statistic since 1993 for the prediction reliability is also represented together with some theoretical models and estimations.

To summarize, one can say that the standard geodetic monitoring [5], the monitoring of different components of the electromagnetic field under, on and over the Earth's surface, some of the atmospheric anomalies and the behaviour of charge distribution in the Earth's radiation belts (see, for example [39] [40]), sometimes can serve as earthquake's precursors.

The progress in the electromagnetic quake earthquake precursor approach was presented in papers [41]-[48]. The earthquake part of the model can be repeated in many time ways "theory-experiment-theory", using nonlinear inverse problem methods for discovering unknown dependences and correlations between different geophysical fields in dynamically changed spaces and times scales.

It seems obvious that the problem of adequate physical understanding of the correlations between electromagnetic precursors, tidal extremes and incoming earthquakes is connected with the progress of creating the adequate Earth's magnetism model.

The achievement of the Earth's surface tidal potential modelling, which includes the ocean and atmosphere tidal influences, is an essential part of the research. In this sense the comparison of the Earth tides codes—ANALYZE, ETERNA-package, the version 3.30 [49]; BAYTAP-G code in the version from November, 15, 1999 [50]-[52]; VAV code (version from April 2002) [53] [54], is very hard but helpful work.

The role of geomagnetic variations as a precursor could be explained by the obvious hypothesis that during the time before the earthquakes, the crust strain, deformation or displacement changes in a certain interval of density volume change, where the chemical phase shift arises, lead to electrical charge shift. The increasing of the electro potential leads to the increasing of the regional Earth currents which influence the geomagnetic field variability. The Fourier analysis of the geomagnetic field gives the time period of alteration in minute scale. We call such a specific geomagnetic variation a geomagnetic quake.

The piezo-effect model for electrical currents cannot explain the alternations due to its linearity.

The K-index [55], accepted for the estimation of the geomagnetic conditions, cannot indicate well the local geomagnetic variation for a minute time period, because it is calculated on the basis of 3-h data. Nevertheless, the K-index behaviour in the near space has to be analyzed because of the possible Sun-wind influence on the local behaviour of the geomagnetic field. If the geomagnetic field components are measured many times per second, one can calculate the frequency dependence of full geomagnetic intensity and analyze the frequency spectrum of the geomagnetic quake. If the variations are bigger than the usual for a certain period of time, one can say that we have a geomagnetic quake, which can be possible earthquake's precursor. The nonlinear inverse problem analysis for 1999-2001 of the geomagnetic and earthquake data for Sofia region gives the estimation that the probability time window for the predicted earthquake (event, events) is approximately ±2 days for the next minimum of the Earth's tidal potential and ±2.7 days for the maximum [47].

The future epicentre coordinates can be estimated according to the data from at least 3 points of measuring the geomagnetic vector, using the inverse problem methods, applied for the estimation of the coordinates of the volume, where the chemical phase shift (dehydration) arrived in the framework of its time window.

In the case of an incoming big earthquake (magnitude > 5 - 6), the changes of vertical electropotential distribution, the Earth's temperature, the infrared Earth's radiation, the behaviour of the water sources, its chemistry and radioactivity (222 Rn), the atmospheric conditions (earthquake clouds, etc.) and the charge density of the Earth's radiation belt, have to be dramatically changed near the epicentre area.

The achievements of the tidal potential modeling of the Earth's surface, including the ocean and atmospheric tidal influences, the multi-component correlation analysis and the nonlinear inverse problem methods in fluid's dynamics and electrodynamics are crucial for every single step of the construction of mathematical and physical models.

## 2. Methodology: The Number Comparison of Two Time Experimental Series

If we need to estimate the variability of one experimental series of measurements' data $T\{T_i, i=1,\cdots,n\}$ we can calculate the mean value $\bar{T}$:

$$\bar{T} = \frac{\sum_{i=1}^{n} T_i}{n} \qquad (1)$$

and to calculate the value of its dimensionless standard deviation *SdT*:

$$SdT = \sqrt{\frac{\sum_{i=1}^{n}\left(1-\frac{T_i}{\bar{T}}\right)^2}{n}} \qquad (2)$$

When one needs to compare the variability of two time experimental series $T_{ij}$, $i = 1, 2$; $j = 1, \cdots, nj$, we have to calculate the its mean value $\bar{T}_i$ and $SdT_i$ according formulae (1) and (2).

Obviously, the biggest standard deviation value means the biggest variability of the series.

## 3. The Forecasting of Imminent Regional Seismic Activity on the Basis of the Geomagnetic and Sun-Moon Tide Code Data

In this paragraph the data-acquisition system for archiving, visualization and analysis is presented [47] [56] [57] [59].

### 3.1. Description of the Approach—Figure 1

The data used:
- the Balkan Intermagnet geomagnetic stations PAG, Bulgaria, SUA, Romania, AQU, Italy, minute data (http://www.intermagnet.org/),
- the software for calculation of the daily and minute Earth tide behaviour (Dennis Milbert, NASA), http://home.comcast.net/~dmilbert/softs/solid.htm),
- the Earth tide extremes (daily average maximum, minimum and inflexed point) as a trigger of earthquakes,
- the data for World A-indices (http://www.swpc.noaa.gov/alerts/a-index.html).

### 3.2. The Variables from Figure 1

The simple mathematics for the calculation of the precursor signal, the software for illustrating the reliability of forecasting and its statistic estimation and the variables in **Figure 1** are described in the next.

The variables $X_m$, $Y_m$, $Z_m$ are the component of minute averaged values of the geomagnetic vector, $m =$ 1440, and the variables $dX_h$, $dY_h$, $dZ_h$ are standard deviation of $X_m$, $Y_m$, $Z_m$, calculated for 1 hour ($h = 1, \cdots, 24$):

$$dX_h = \sqrt{\frac{\sum_{m=1}^{60}\left(1 - \frac{X_m}{\overline{X}_h}\right)^2}{60}}, \qquad (3)$$

where:

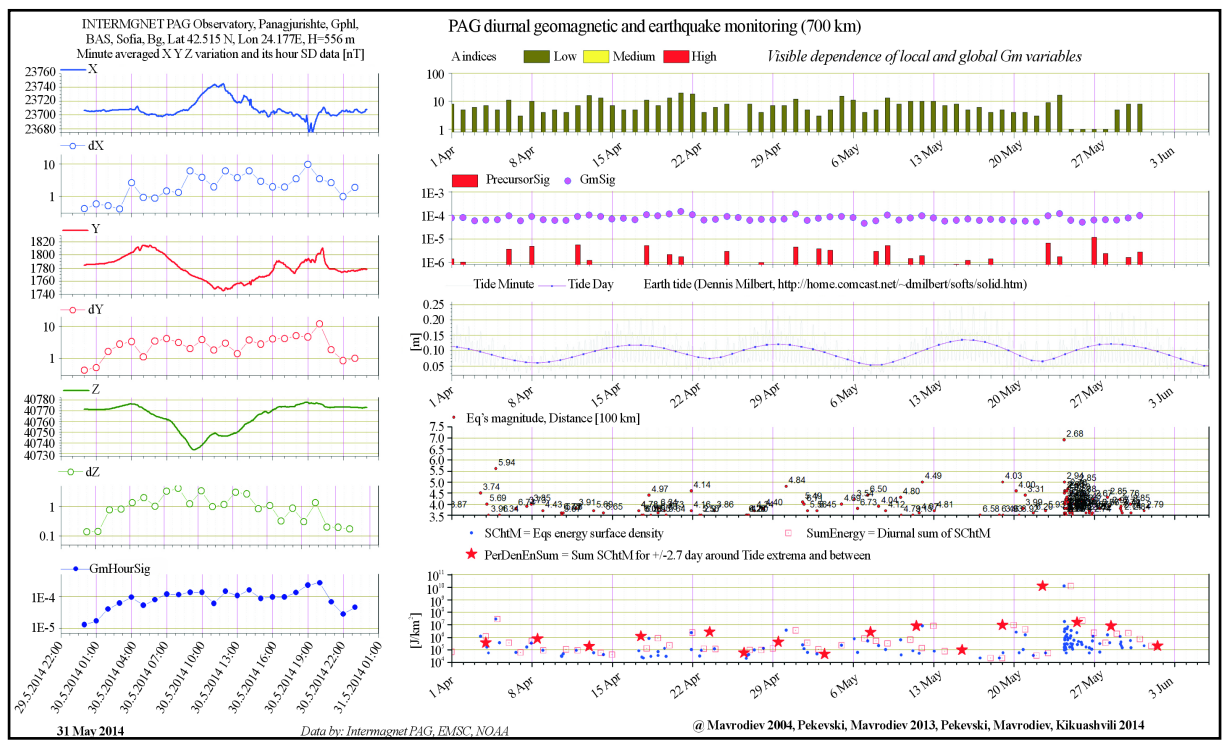

**Figure 1.** Reliability of the approach for the PAG INTERMAGNET station located in Bulgaria.

$$\bar{X}_h = \frac{\sum_{m=1}^{60} X_m}{60}. \quad (4)$$

The geomagnetic signal $\text{GeomHourSig}_h$ is:

$$\text{GeomHourSig}_h = \sqrt{\frac{dX_h^2 + dY_h^2 + dZ_h^2}{X_h^2 + Y_h^2 + Z_h^2}}. \quad (5)$$

The A indices are the Low, Medium and High a-indices calculated by the NOAA, Space weather prediction center: http://www.swpc.noaa.gov/alerts/a-index.html. In the approach we use the Alow;

The variable $\text{GmSig}_{day}$ is the diurnal mean value of GmHourSigh:

$$\text{GeomSig}_{day} = \frac{\sum_{h=1}^{24} \text{GeomHourSig}_h}{24} \quad (6)$$

and $\text{PrecursorSig}_{day}$

$$\text{PrecursorSig}_{day} = 2 \frac{\text{GeomSig}_{day} - \text{GeomSig}_{yesterday}}{\text{Alow}_{day} + \text{Alow}_{yesterday}} \quad (7)$$

The indices of earthquake's magnitude value are the distance in hundred km between the epicentre and the monitoring point.

The variable *SChtM* is the modified earthquake's surface energy density in the monitoring point:

$$SChtM = \frac{10^{(1.4M+4.8)}}{(40+\text{Depth}+\text{Distance})^2} \left[\text{J}/\text{km}^2\right]. \quad (8)$$

The variable PerDenEqSum [J/km²] is the sum of variable *SChtM* for all earthquakes occurred in the time period ±2.7 days before and after the tide extreme. Obviously its value can serve as estimation of the regional seismic activities for the time period around the tide's extreme;

The variable SumEnergy [J/km² per day] is the sum of the variable *SChtM* for all earthquakes occurred during the day in the 700 km region. This variable can serve as adequate diurnal quantitative measure of regional seismicity;

One has to note that the explicit form of variable *SChtM* was established in the framework of inverse problem with condition to have more clear correlation between the variables $\text{PrecursorSig}_{day}$ and PerDenEqSum.

The variable *TideMinute* [cm] is the module of tide vector calculated every 15 minutes;

The variable *TideDay* [cm] is the diurnal mean value in time calculated in the analogy of mass centre formulae

$$\text{Time}_{TideDay} = \frac{\sum_{m=1}^{360} m \text{TideDay}_m}{\sum_{m=1}^{360} \text{TideDay}_m}. \quad (9)$$

Note: for seconds and more samples per second the generalization has to calculate geomagnetic field characteristics for every minute and correspondingly the values of GmSigday have to be the average for 1440 minutes.

The positive value of variable $\text{PrecursorSig}_{day}$ (Equation (6)) means that the geomagnetic field variability, which is calculated by relative standard deviations of geomagnetic field components, is increasing ((Formulae (1), (2), (3)). In analogy with earthquake we define such increasing as *geomagnetic quake*. As one can see from **Figure 1**, the appearance of geomagnetic quake shows the regional seismic activity increase (the bigger value of the PerDenEqSum variable) in the time period of the further tide extreme. So, the described geomagnetic quake approach can serve as forecasting for imminent regional seismic activity.

In the **Figure 1** the values of variable PerDenEqSum are calculated not only in the time periods around extremes but also in the time period between them. It is seen that the values in the extreme periods are bigger.

The use of above described analysis for longer time period with calculation of distribution of day difference between "predicted" earthquakes (earthquakes with biggest value *SChtM*) in the frame work of special created data acquisition system demonstrates the reliability of the approach for forecasting of imminent regional seismic activity.

## 4. The Reliability of the Approach Using INTERMAGNET Data from Stations Located in Bulgaria, Romania and Italy

The distribution of differences between the times of the "predicted" and occurred earthquakes can serve as statistical proof for the reliability of the forecasting of imminent regional seismic activity. As a definition of a "predicted" earthquake, we use the natural hypothesis that it is the earthquake with biggest value of variable among all the earthquakes occurred in the region with a radius of 700 km. The distribution of day differences is presented for Intermagnet data stations PAG, Panagurishte, Bulgaria, SUA, Surlari, Romania and AQU, L'Aquila, Italia. The corresponding data parameters and time periods are presented in the **Figures 2-4**.

As one can see from the **Figure 2**, the distribution is close to the Gauss one with $hi^2 = 0.89$. The relation between the sums of *SChtM* for the predicted and occurred earthquakes is equal to $r = 5.51/6.48$. So, the reliability of regional imminent seismic forecasting has 85% probability for the PAG data.

As one can see from the **Figure 3**, the distribution is close to the Gauss one with $\chi^2 = 0.92$. The relation between the sums of *SChtM* for the predicted and occurred earthquakes is equal to $r = 8.83/11.0$. So, the reliability of regional imminent seismic forecasting has 80% probability for the SUA data.

As one can see from the **Figure 4**, the distribution is close to the Gauss one with $hi^2 = 0.87$. The relation between the sums of *SChtM* for the predicted and occurred earthquakes is equal to $r = 1.05/1.15$. So, the reliability of regional imminent seismic forecasting has 91 % probability for the AQU data.

## 5. Some Examples of Imminent Regional Confirmation of Forecasting

Some examples (http://theo.inrne.bas.bg/~mavrodi):
- Dusheti, Georgia flux gate second magnetometer—Mw 7.1, depth 7.2 km, 2011, 23 Oct, 36.63N, 43.49E, Van, Turkey earthquake;

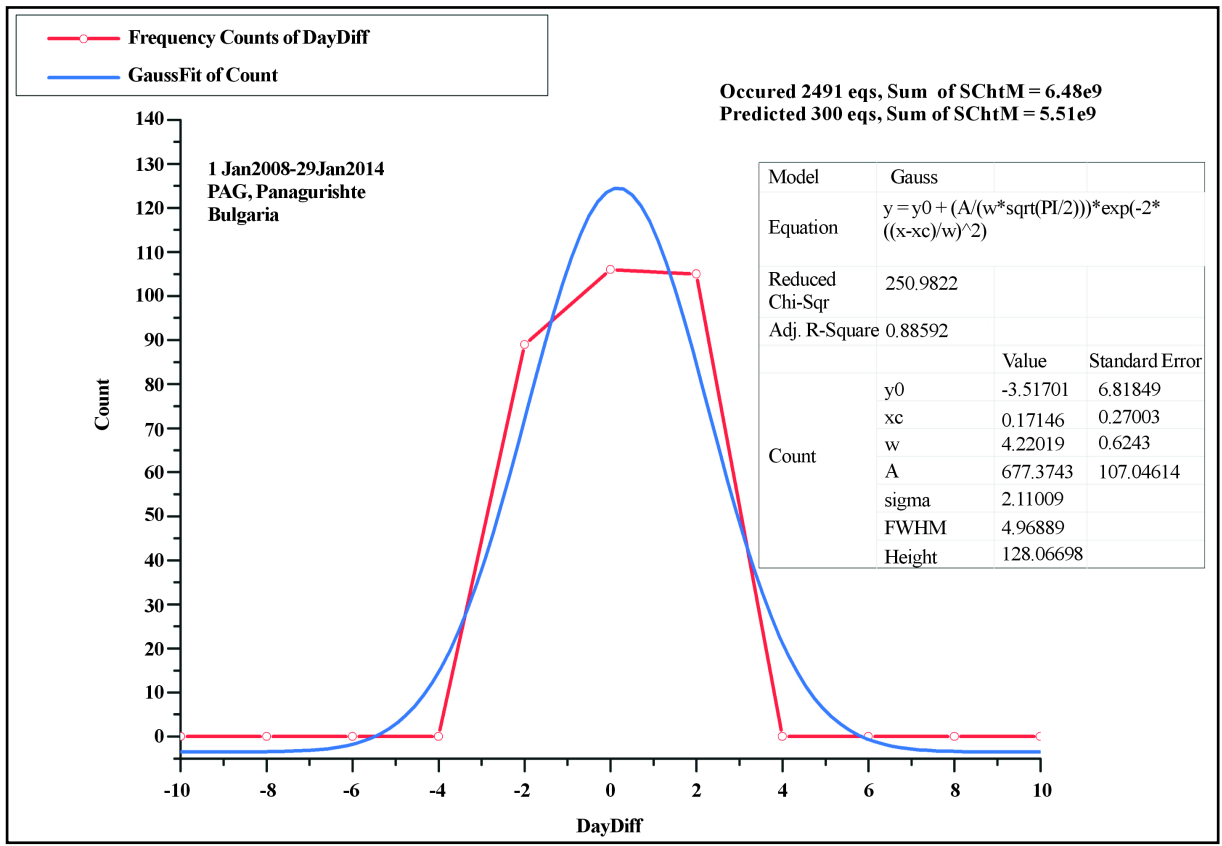

**Figure 2.** Day difference distribution for PAG data and it Gauss fit.

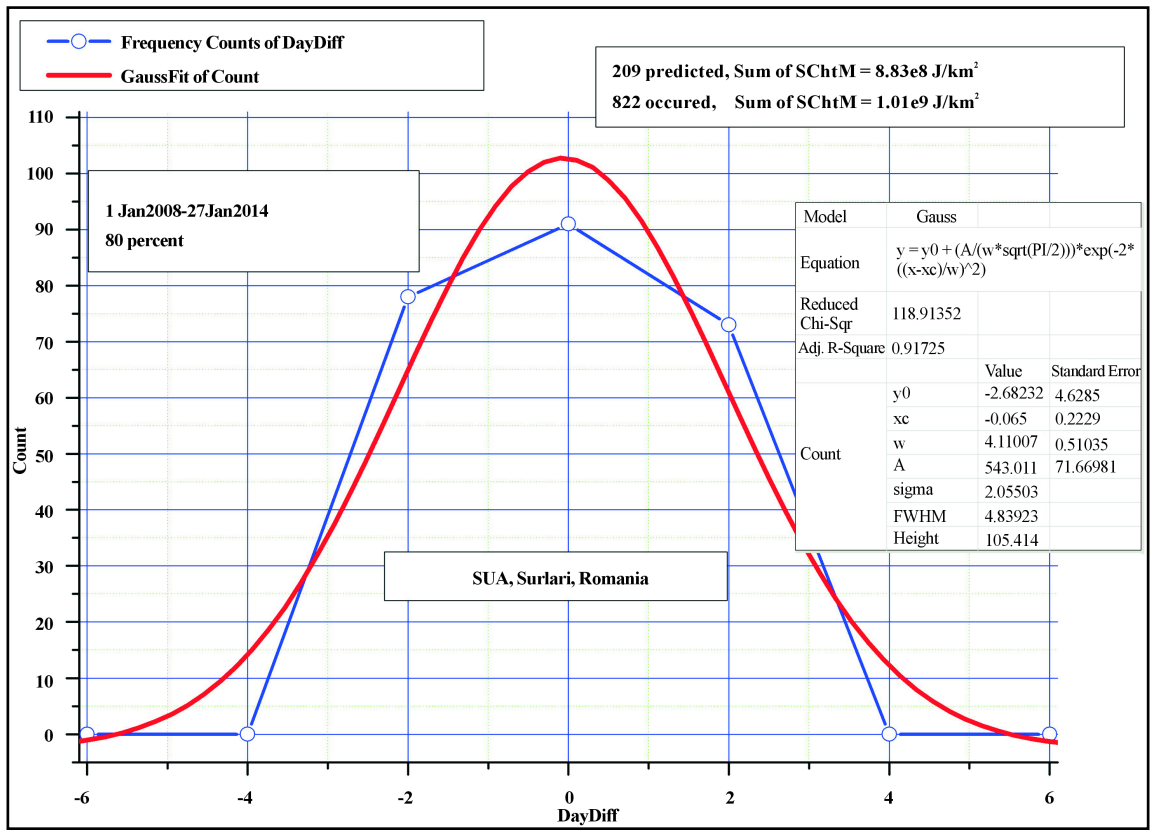

**Figure 3.** Day difference distribution for SUA (Surlari, Romania) data and its Gauss fit.

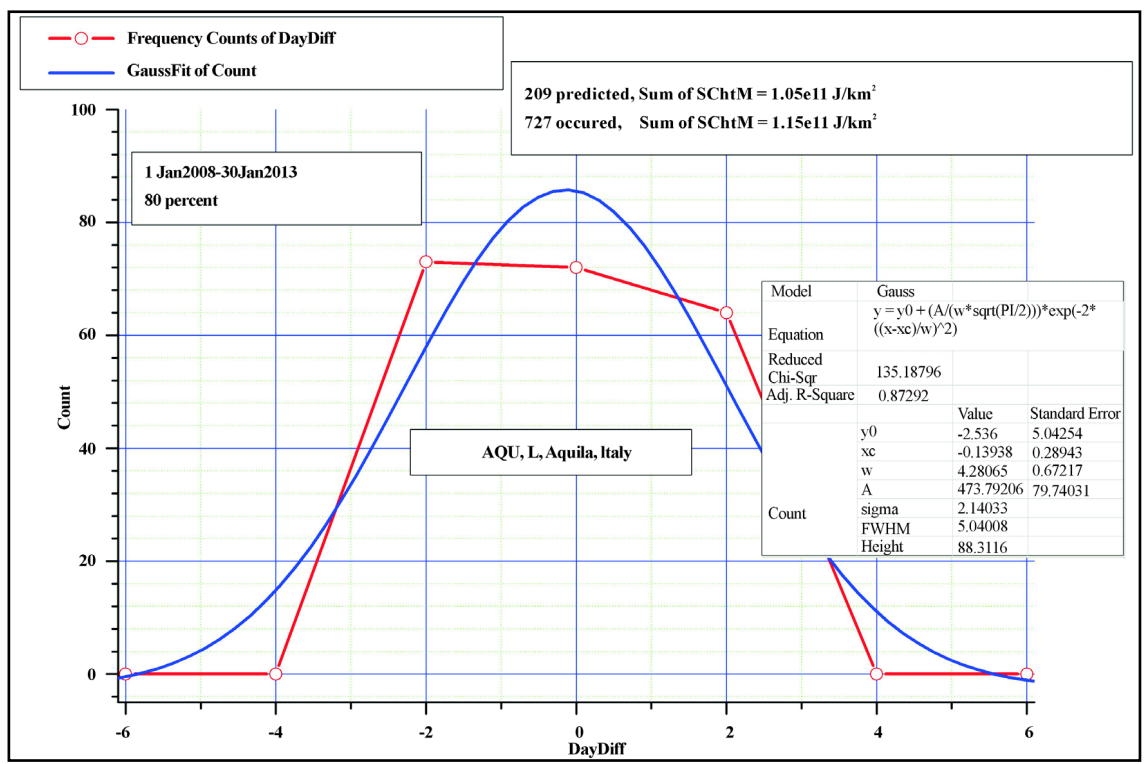

**Figure 4.** Day difference distribution for AQU (L'Aquila Italy) data and its Gauss fit.

- Skopje, Macedonia (second) and Panagurichte (minute) flux gate magnetometers—Mw 5.6, Depth 9.4 km, 42.66N, 23.01E, 00.00 hour, 22 May, 2012;
- Grocka, Serbia and Panagurichte (minute) flux gate magnetometers—Mw 6.1, Depth 18 km, 26 Jan 2014, 13:55, 38.19N, 20.41E; Mw 6.0, depth 2 km, 3 Feb, 2014, 03:08, 38.25N, 20.32E, Mw 5.6, Depth 9.4 km.

During our investigation of the relation between the regional geomagnetic and seismic activity in the areas of interest it was found that in case of a strong earthquake occurred in the distances less than 700 km from the geomagnetic observatory, a clear precursor signal was evident.

## 6. Using the Daily Tide Extremes as Earthquakes Triggers

In this section we present an explanation why the geomagnetic quakes forecasting are reliable.

We introduce the variable **DayDiff** which is the difference between the time of earthquake and the time of nearest tide extreme:

$$\text{DayDiff}_i = \text{TimePredicted}Eq_i - \text{TimeOcurred}Eq_i, [\text{Day}], \, i = 1,\cdots,$$

Number of Earthquakes and $\text{TimePredicted}Eq_i$ is the time of nearest tide extreme.

The distribution for 628,873 world earthquakes with magnitude M ≥ 3.5 of occurred since 1981 (International Seismological Center: http://www.isc.ac.uk/data) of variable **DayDiff** is presented in **Figure 5**. As one can see 92% of all analyzed earthquakes are occurred in the time period ±2.23 day around the time of tide extreme in their epicentres.

## 7. Proposal for Complex Program for Solving "When, Where and How" Earthquake's Prediction Problem

For the solving step by step "when, where and how" earthquake prediction problem we need realization of project ([58] [59] (http://theo.inrne.bas.bg/~mavrodi/project.html)), [60]:

Step 1. Creating complex regional monitoring network with a real time data acquisition system for archiving, testing, visualizing and analyzing the data and the estimation risks, which includes: 3 stationary and one mobile geoelectromagnetic stations (1 - 10 samples per second), 3 - 5 water level boreholes (2 samples per hour), 5 - 7 stations for measure Radon (222 Rn) concentration (6 samples per day), the seasonal and independent daily data for the crust temperature (2 samples per day) and 3 - 4 multi-frequency sound vector receivers (6 samples per hour). Important note: the number of samples is preliminary estimated.

Step 2. Using the above data, the regional geological and seismic (especially foreshock) data, the ionosphere behaviour data to start formulating and solving the inverse problem for forecasting the time, the epicentre's coordinates, the depth and the intensity of an incoming regional earthquake as well as for actualization of regional seismic risk estimation.

Such complex monitoring polygons can be created in Balkans, Ukraine and Caucasus for better practical using of EU IRSES BlackSeaHazNet project (2011-2014) results and qualified staff: BlackSeaHazNet results: http://cordis.europa.eu/result/rcn/148955_en.html [60].

## 8. Conclusions

The correlation between local geomagnetic quakes and incoming earthquakes in the time window defined from the further tide extreme (approximately ±2.7 days) is tested statistically. The distribution of the time difference between the predicted and occurred events has a Gauss fit with adjusted R-square approximately equal to 0.89 and *w* = 4.2 ± 0.6.

This result can be interpreted like reliable approach for solving the "when" regional imminent seismic activity problem.

The discrepancy between values of Gauss fit parameter **w** for different cases is probably connected with INTERMAGNET one-minute samples.

## Acknowledgements

The results presented in this paper rely on the data collected at INTERMAGNET PAG (Panagjurishte, Bulgaria), SUA (Surlari, Romania), AQU (L'Aquila, Italy) stations. We thank the Geophysical Institute of the Bulgarian

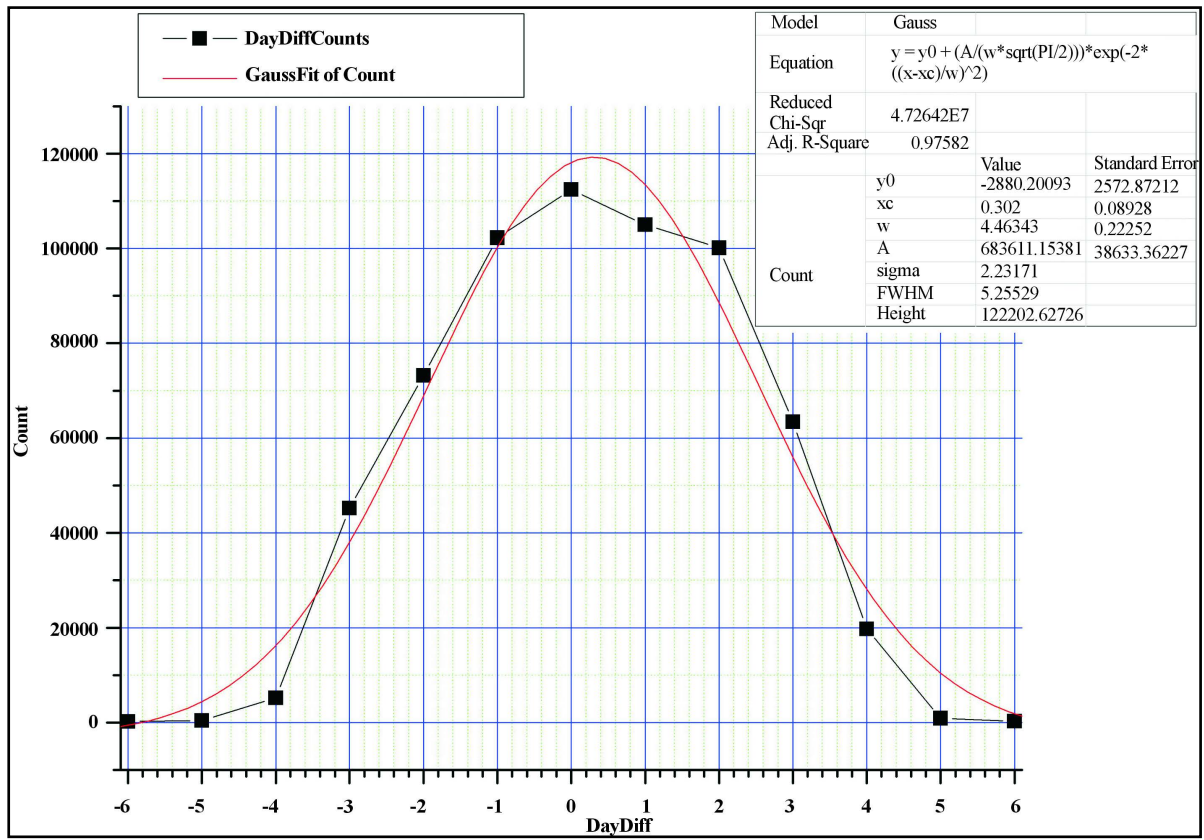

**Figure 5.** Day difference distribution for AQU (L'Aquila Italy) data and its Gauss fit.

Academy of Science for supporting PAG operation and INTERMAGNET for promoting high standards of magnetic observatory practice (www.intermagnet.org).

The financial support in the framework of FP7, Marie Curie Actions, International Research Staff Exchange Scheme, Project title: Complex Research of Earthquake's Forecasting Possibilities, Seismicity and Climate Change Correlations, Acronym: BlackSeaHazNet, Grant Agreement Number: PIRSES-GA-2009-246874 as well the support of INRNE, BAS, are highly appreciated.

We would like to thank heartily the REA project officers heartily Dr. Oscar Perez-Punzano and Dr. Atantza Uriarte-Iraola, for their invaluable support in the process of negotiation and executing the project as well as to express our sincere gratitude to the DG Institute for Nuclear Research and Nuclear Energy, BAS in the face of Directore Assoc. Prof. Dr. Dimitar Tonev and Vice Director Assoc. Prof. Dr. Lachezar Georgiev, for the encouragement and valuable assistance, to Dr. Frank Marx, and in his face—to the European Commission, for the financial support for the BlackSeaHazNet project realization.

We would like to note the technical help of RadkaMavrodieva and DimitarHaydutov.

## Conflict of Interest

The authors declare that they have no conflict of interest.